 \documentclass[preprint]{aastex}

\usepackage{psfig}
\newbox\grsign \setbox\grsign=\hbox{$>$} \newdimen\grdimen
\grdimen=\ht\grsign
\newbox\simlessbox \newbox\simgreatbox \newbox\simpropbox
\setbox\simgreatbox=\hbox{\raise.5ex\hbox{$>$}\llap
     {\lower.5ex\hbox{$\sim$}}}\ht1=\grdimen\dp1=0pt
\setbox\simlessbox=\hbox{\raise.5ex\hbox{$<$}\llap
     {\lower.5ex\hbox{$\sim$}}}\ht2=\grdimen\dp2=0pt
\setbox\simpropbox=\hbox{\raise.5ex\hbox{$\propto$}\llap
     {\lower.5ex\hbox{$\sim$}}}\ht2=\grdimen\dp2=0pt
\def\simgreat{\mathrel{\copy\simgreatbox}}
\def\simless{\mathrel{\copy\simlessbox}}

\shorttitle{An \emph{XMM-Newton} Study of 30 Dor C} 
\shortauthors{Smith and Wang}

\begin{document}

\title{Confronting the Superbubble Model with X-ray Observations of 30
Dor C}


\author{David A. Smith and Q. Daniel Wang}

\affil{Department of Astronomy, University of Massachusetts, Amherst,
MA 01003-9305; dasmith@xray.astro.umass.edu, wqd@astro.umass.edu}


\begin{abstract}

We present an analysis of \emph{XMM-Newton} observations of the
superbubble 30 Dor C and compare the results with the predictions from
the standard wind-blown bubble model.  We find that the observed X-ray
spectra cannot be fitted satisfactorily with the model alone and that
there is evidence for nonthermal X-ray emission, which is particularly
important at $\simgreat 4$~keV.  The combination of the bubble model
and a power-law gives a reasonable fit to the observed spectra.  The
thermal pressure and central temperature of the bubble are $3.3 \times
10^{-11}$ dyne cm$^{-2}$ and $7.4 \times 10^{6}$~K, respectively, and
we infer that, for a bubble age of $t \sim 4 \times 10^{6}$ years, the
ambient density is $n_{0} \simeq 38$ cm$^{-3}$, the mechanical
luminosity is $L_{\rm mech} \sim 10^{37}$ erg s$^{-1}$, and the
coefficient of thermal conductivity is $\sim 0.05$ of the Spitzer
value.  The total unabsorbed $0.1$--$10$~keV luminosities of the
eastern and western parts of the bubble are $\simeq 3 \times 10^{36}$
erg s$^{-1}$ and $\simeq 5 \times 10^{36}$ erg s$^{-1}$, respectively.
The unabsorbed $0.1$--$10$~keV luminosity of the bubble model is $\sim
4 \times 10^{36}$ erg s$^{-1}$ and so the power-law component
contributes between $1/3$ and $1/2$ to the total unabsorbed luminosity
in this energy band. The nature of the hard nonthermal emission is not
clear, although recent supernovae in the bubble may be responsible. We
expect that about one or two core-collapse supernovae could have
occured and are required to explain the enrichment of the hot gas, as
evidenced by the overabundance of $\alpha$-elements by a factor of
$\gtrsim 3$, compared to the mean value of $\sim 0.5$ solar for the
interstellar medium in the Large Magellanic Cloud.  As in previous
studies of various superbubbles, the amount of energy currently
present in 30 Dor C is significantly less than the expected energy
input from the enclosed massive stars over their lifetime.  We
speculate that a substantial fraction of the input energy may be
radiated in far-infrared by dust grains, which are mixed with the hot
gas because of the thermal conduction and/or dynamic mixing.
 
\end{abstract}


\section{INTRODUCTION}

The fast winds from massive stars and their subsequent supernovae
(SNe) ejecta dominate the mass and energy input to the interstellar
medium (ISM; e.g., Abbott 1982) and so may regulate star formation and
galaxy evolution.  Most massive stars form in OB associations, and the
concentrated energy input from massive stars in OB associations sweeps
up the ambient ISM to form expanding shells called superbubbles (e.g.,
McCray \& Snow 1979).  The physical structure of a superbubble is very
similar to that of a bubble blown by the stellar wind of an isolated
massive star \citep{mm88}.  However, theories of such bubbles are
still very uncertain and have not had much serious appraisal with
observations of high quality.

Studies of Galactic superbubbles are difficult because of their large
angular sizes and the confusion of objects in the line of sight.  The
Large Magellanic Cloud (LMC), in contrast, provides an excellent
laboratory to study superbubbles since it is nearby (at a distance of
$50$~kpc, as adopted here, $1^{\prime\prime} = 0.24$ pc; Eastman \&
Kirshner 1989) and almost face-on (inclined by $\sim 35^{\circ}$ to
the line of sight; van der Marel \& Cioni 2001). It also has little
foreground ($E_{\rm B-V} = 0.08 \, \rm mag$; Xu et al. 1992) or
internal extinction ($E_{\rm B-V}$ in the range from $\sim 0.06$ to
$\sim 0.3\, \rm mag$; Bessel 1991).  The superbubble 30 Dor C (= DEM L
263; Davies, Elliott, \& Meaburn 1976) is southwest of the main 30
Doradus complex in the LMC, and is coincident with the large OB
association LH 90 (= NGC 2044; Lucke \& Hodge 1970).  This OB
association consists of a number of stellar clusters and is
particularly rich in Wolf-Rayet (W-R) stars \citep{lt84}.  The most
massive stars appear to have ages in the range from $\sim 3$ to $\sim
7$~Myr (Testor, Schild, \& Lortet 1993).  In X-ray imaging
observations, 30 Dor C appears almost as a complete shell with a
diameter of $\sim 7^{\prime}$ (Wang \& Helfand 1991; Dennerl et
al. 2001; Dunne, Points, \& Chu 2001).  The X-ray shell is confined
within the H$\alpha$-emitting filaments (see Fig.~1\emph{l} of Dunne
et al. 2001), as expected from the evaporation of cool gas into the
hot interior of a superbubble.  There is also an indication for a
substantial nonthermal contribution in addition to an assumed
one-temperature thermal component, particularly in the western part of
shell \citep{bam03,bam04}.  The radio emission from 30 Dor C may be
due to a combination of synchrotron-emitting relativistic electrons
produced by shocks within the superbubble and thermal emission from
the H$\alpha$-emitting shell \citep{mat85}.

In this paper, we report on a spatially resolved spectroscopic
analysis with \emph{XMM-Newton} of the diffuse X-ray emission
emanating from 30 Dor C.  We specifically use the wind-blown bubble
model of Weaver et al. (1977; see also Mac Low \& McCray 1988) to
infer the temperature and density structure of the X-ray emitting gas,
which in turn are used to estimate both the density of the ambient
medium and the rate at which energy is injected into the ISM by the
stellar winds.  

\section{OBSERVATIONS AND DATA REDUCTION}
\label{sec-obs}

The region surrounding 30 Dor C is known to be rich in X-ray sources
and was chosen as the first light image for \emph{XMM-Newton}
\citep{jan01}.  However, data acquired this early in the mission are
in a format which cannot be processed by the Science Survey Center
using the standard software and so are not included in the
\emph{XMM-Newton} archive at the time of writing.  The data reported
here were instead acquired during $\simeq 21.6$ and $\sim 37.5$~ks
observations of SN 1987A and PSR J0537-6909 with \emph{XMM-Newton} on
2000 November 25-26 (observation ID 0104660301; PI: Watson) and 2001
November 19-20 (observation ID 0113020201; PI: Aschenbach),
respectively.  These observations had a low background count rate and
were pointed $\sim 6^{\prime}$ (0104660301) and $\sim 9^{\prime}$
(0113020201) offset from the center of the 30 Dor C.  While there are
other observations that covered 30 Dor C in the \emph{XMM-Newton}
archive, these observations are either severely contaminated by
background flares (observational ID 0083250101) or had a very short
good exposure time (e.g., $\sim 3.3$~ks for observation 0104660101;
PI: Watson).  The EPIC MOS cameras \citep{tur01} were operated in the
full-frame mode with the medium filter inserted.  Over the time
interval between the two observations, the energy resolution of MOS
CCDs had worsened, from $140$~eV to $150$~eV (FWHM) at $5.9$~keV, due
to radiation damage.  These values are actually worse than the initial
post-launch value of $\sim 130$~eV.  The on-axis telescope point
spread function is $4.\!^{\prime\prime}3$ (FWHM) at 1.5~keV and the
field-of-view is roughly circular with a diameter of $\sim
30^{\prime}$.  Unfortunately, the EPIC pn exposure of the first
observation was taken with the calibration lamps switched on, so
rendering the data useless for scientific analysis, and the pn was
operated in the timing mode during the second observation, so
providing no imaging data.

The data reduction and analysis were done using the XMM-Newton Science
Analysis Software (SAS) version 5.4.1 (released on 2003 January 16).
The MOS data were reprocessed into calibrated event list files using
the SAS program \emph{emchain}.  The procedure of \citet{rp03} was
used to filter each event list for periods of high background.  We
created lightcurves in the 10--15 keV energy range, where no source
counts are expected due to the very low effective area of the X-ray
telescope, and using only single pixel events and flag values as
defined by \#XMMEA\_EM.  Periods when the count rate in these
lightcurves exceeded $0.35$ counts s$^{-1}$ were excluded from the
event files.  The event lists were then filtered further, keeping only
single, double, triple, and quadruple pixel events in the
$0.2$--$15$~keV energy range.  For the first observation, the total
good exposure time was $\simeq 21$~ks, and for the second observation,
the MOS1 and MOS2 CCDs had total good exposure times of $\simeq 17$~ks
and $\simeq 13.5$~ks, respectively.  We also checked for soft proton
flares in the $2$--$8$~keV band, but found none.

\begin{figure} 
\centerline{ 
\psfig{figure=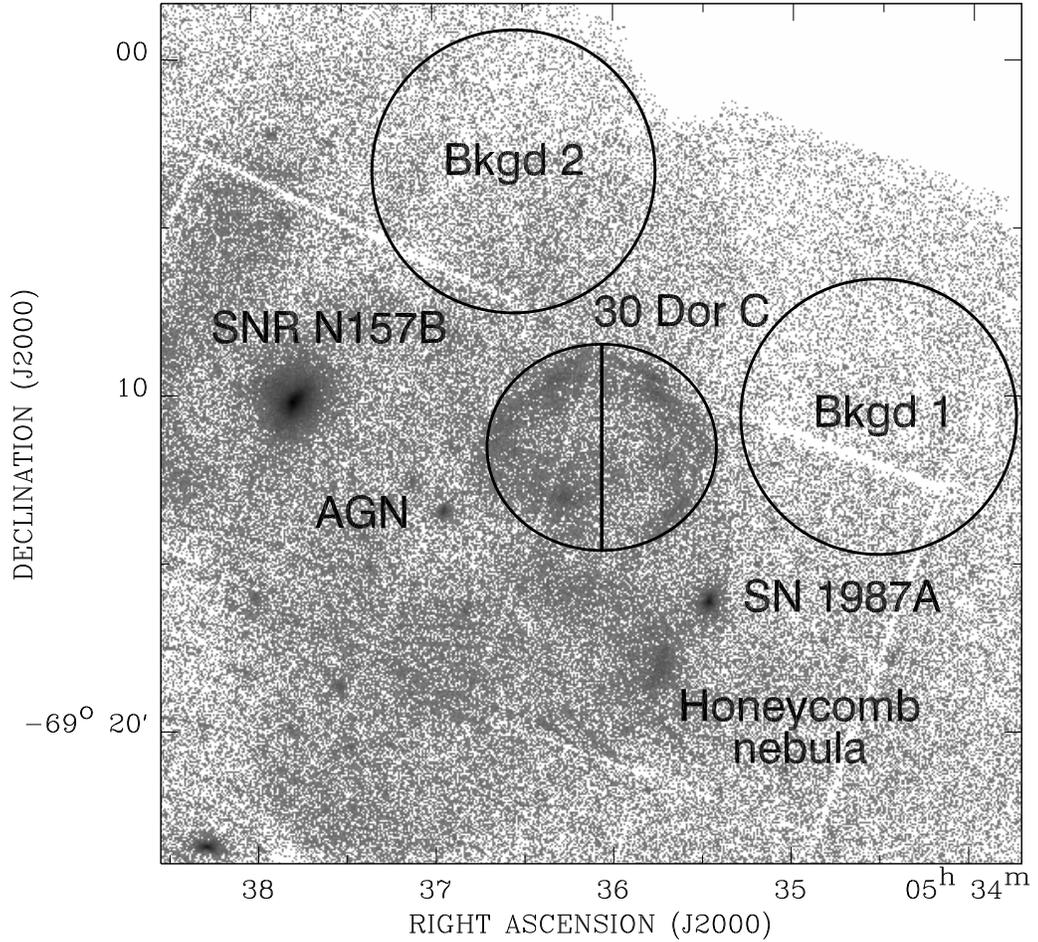,height=6in,angle=90,clip=}
}
\caption{The \emph{XMM-Newton} image of the 30 Dor C field in the
$0.3$--$8$~keV band with the bright sources of X-ray emission
labelled.  The image has a pixel size of $3^{\prime\prime} \times
3^{\prime\prime}$ ($0.72 \, \rm pc \times 0.72 \, \rm pc$).  The solid
black lines mark source and background (Bkgd 1 for the first
observation; Bkgd 2 for the second observation) regions from which
spectra were extracted.  The shading is proportional to the logarithm
of the intensity and ranges from $10^{-3}$ (white) to $830$ (black)
counts pixel$^{-1}$.}
\label{fig1}
\end{figure}

The interstellar extinction ($E_{\rm B-V}$) towards the western part
of 30 Dor C is known to be higher by at least $0.4 \, \rm mag$ than
that towards the eastern part (Testor et al. 1993; see also Dunne et
al. 2001 and Fig.~5 of Dennerl et al. 2001).  We therefore extracted
spectra from two sectors of an ellipse centered on 30 Dor C with
major-axis diameter of $\simeq 6.\!^{\prime}9$ ($\simeq 100$~pc) and
major-to-minor axis ratio of $0.9$.  The regions are marked in
Figure~\ref{fig1}, which is an image of the central region of the 30
Dor C field (see \S~\ref{sec-morphology} for a description of X-ray
sources in this image).  The shape and size of the extraction regions
were chosen to include as much of the diffuse emission from the
superbubble as possible, while minimizing the contribution from the
background.

The background was estimated from source free regions near the
superbubble (Fig.~\ref{fig1}).  There are spatial variations in the
detector background rate, which can be estimated from a comparison of
blank-field spectra in the source and source free regions of the 30
Dor C observations.  For this comparison, we have used blank-field
observations with a total exposure of $\simeq 488$~ks and $\simeq
593$~ks for the MOS1 and MOS2 detectors, respectively \citep{rp03}.
These data, taken with the same instrument mode and filter combination
as the 30 Dor C observations, had point sources of emission excised,
and were reprojected on the sky to match the positions of the 30 Dor C
observations.  

We find that for the first observation, the MOS1 blank-field spectra
in the source free region and in the eastern part of 30 Dor C are
statistically consistent (within $3\sigma$ deviation) with each other
in the $0.3$--$10$ keV band, with $\chi^{2} = 573.5$ (497 dof).
However, for the same observation, a comparison of the MOS2
blank-field spectra in the source free region and in the eastern part
of 30 Dor C revealed significant differences in the background count
rate near the energy of the Si K$\alpha$ ($1.7$~keV) fluorescence
line.  We therefore compared the MOS2 blank-field spectra in the
$0.3$--$1.6$ and $1.85$--$10$ keV band, and find that the spectra are
consistent with each other, with $\chi^{2} = 510.4$ (466 dof).  There
are also, for the first observation, and near the energies of the Al
K$\alpha$ (1.5 keV) and the Si K$\alpha$ fluorescence lines,
significant differences in the background count rate between the
blank-field spectra in the source free region and the blank-field
spectra in the western part of 30 Dor C.  In the $0.3$--$1.3$ and
$1.85$--$10$ keV band, however, the blank-field spectra in the source
free region are statistically consistent with the blank-field spectra
in the western part of 30 Dor C, with $\chi^{2} = 465.6$ (449 dof) and
$\chi^{2} = 458.6$ (430 dof) for the MOS1 and MOS2 detectors,
respectively.

Similarly, for the second observation, and near the energy of the Si
K$\alpha$ fluorescence line, we find that there are significant
differences in the background count rate near between the blank-field
spectra in the source free region and the blank-field spectra in the
eastern part of 30 Dor C.  In the $0.3$--$1.6$ and $1.85$--$10$~keV
band, however, the blank-field spectra are consistent with each other,
with $\chi^{2} =501.4$ (464 dof) and $\chi^{2} = 471.0$ (476 dof) for
the MOS1 and MOS2 detectors, respectively.  There are also, for the
same observation, and near the energies of the Al K$\alpha$ and Si
K$\alpha$ fluorescence lines, significant differences in the
background count rate between the blank-field spectra in the source
free region and the blank-field spectra in the western part of 30 Dor
C.  We therefore compared the blank-field spectra in the $0.3$--$1.3$
and $1.85$--$10$ keV band, and find that the MOS1 and MOS2 blank-field
spectra in the source free region should be scaled by $0.92$ and
$0.84$, respectively, in order to match the blank-field spectra in the
western part of 30 Dor C.  The rescaled spectra are statistically
consistent with the blank-field spectra in the western part of 30 Dor
C, with $\chi^{2} = 546.3$ (478 dof) and $\chi^{2} = 451.1$ (468 dof)
for the MOS1 and MOS2 detectors, respectively.

We therefore excluded data in the $1.6$--$1.85$ keV band from the MOS2
spectra of the diffuse emission in the eastern part of 30 Dor C and
data in the $1.3$--$1.85$ keV band from the spectra of the diffuse
emission in the western part of 30 Dor C.  For the second observation,
we also excluded data in the $1.6$--$1.85$ from the MOS1 spectrum of
the diffuse emission in the eastern part of 30 Dor C.

We created spectral redistribution matrices and ancilliary response
files for the source spectra using the SAS programs \emph{rmfgen} and
\emph{arfgen}, respectively.  The ancilliary response file appropriate
for each source spectrum was created weighting the position-dependent
quantum efficiency and effective area by the number of $0.3$--$10$~keV
blank-field background subtracted counts in the corresponding region.
Prior to performing the spectral analysis with XSPEC version 11.2.0au
\citep{arn96}, we rebinned the data so that the background subtracted
signal-to-noise ratio in each bin is greater than $4$.  When
performing the spectral analysis, we excluded data with energies
greater than $10$~keV, since the background dominates the total
emission from 30 Dor C at energies greater than this.  We also
excluded data with energies less than $0.3$~keV, because of the
uncertainties in the calibration of the instruments.

\section{ANALYSIS AND RESULTS}
\label{sec-analysis}

\subsection{Morphological Properties} 
\label{sec-morphology}

Figure~1 shows the \emph{XMM-Newton} image of the $\sim 26^{\prime}
\times 26^{\prime}$ ($370 \, \rm pc \times 370 \, \rm pc$) region
centered on 30 Dor C in the $0.3$--$8$~keV band.  In this image, 30
Dor C is a roughly elliptical feature with major axis $\sim
7^{\prime}$ ($100$~pc).  This feature appears to be the observational
manifestation of a bubble blown in the ISM by the winds from massive
stars in the bubble's interior \citep{mm88}.  Other sources of X-ray
emission in Figure~\ref{fig1} include SN 1987A, which is located $\sim
6^{\prime}$ southwest of the bubble center and close to a region of
diffuse X-ray emission associated with the Honeycomb nebula
\citep{mea93,chu95b}.  There is also a compact X-ray source,
J0536.9-6913, to the east of 30 Dor C, which has a radio counterpart
and is probably a background AGN \citep{hab01}.  The bright source at
the left of the image is the Crab-like supernova remnant N157B
\citep{wg98}.

\begin{figure} 
\centerline{ 
\psfig{figure=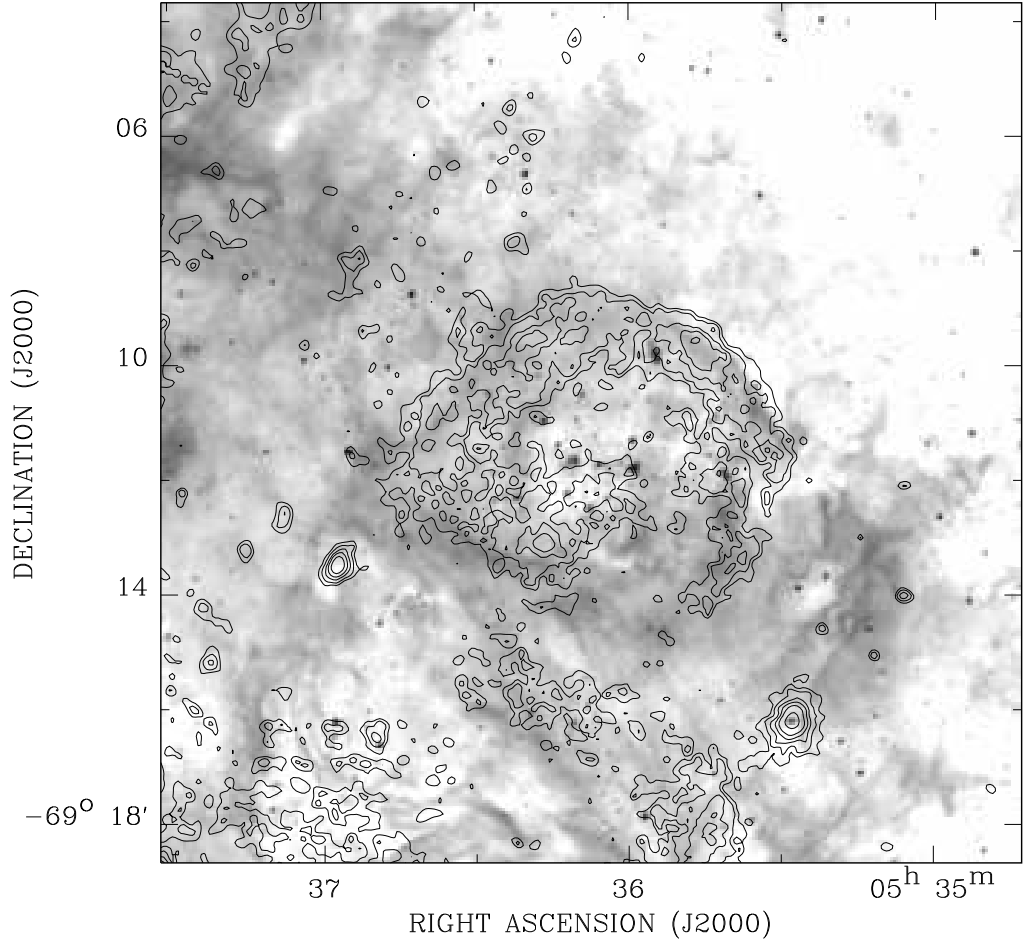,height=6in,angle=90,clip=}
}
\caption{An adaptively smoothed \emph{XMM-Newton} image (contours) of
30 Dor C in the $0.3$--$8$~keV band superposed on an H$\alpha$ MCELS
image (grayscale) at $6563$\AA \, (Smith et al. 1998).  In the
adaptive smoothing process, exposure corrected images are convolved
with a two-dimensional Gaussian, with its width adjusted to achieve a
count-to-noise ratio greater than 6.  The shading is proportional to
the logarithm of the intensity and ranges from $57$ (white) counts
pixel$^{-1}$ to $4200$ (black) counts pixel$^{-1}$ in the optical
image.  Countours are drawn at $(1.4, 1.9, 3.0, 4.8, 8.9$, and $33)
\times 10^{-2}$ counts s$^{-1}$ arcmin$^{-2}$ in the X-ray image.}
\label{fig2}
\end{figure}

In Figure~\ref{fig2}, we compare an H$\alpha$ emission-line image from
University of Michigan/CTIO Magellanic Cloud Emission-line Survey
(MCELS; Smith et al. 1998) to an adaptively smoothed image of the
inner $15^{\prime} \times 15^{\prime}$ region of the 30 Dor C field in
the $0.3$--$8$~keV band.  The edge-brightened X-ray emission is
presumably due to an increase in gas density around the edge of the
bubble, and is spatially enclosed by the H$\alpha$-emitting filaments.
It is conceivable that the increase in gas density is due to the
evaporation of cool ($\sim 10^{4}$~K) gas into the hot interior from
the surrounding H$\alpha$-emitting shell \citep{wea77}.

\begin{figure} 
\centerline{ 
\psfig{figure=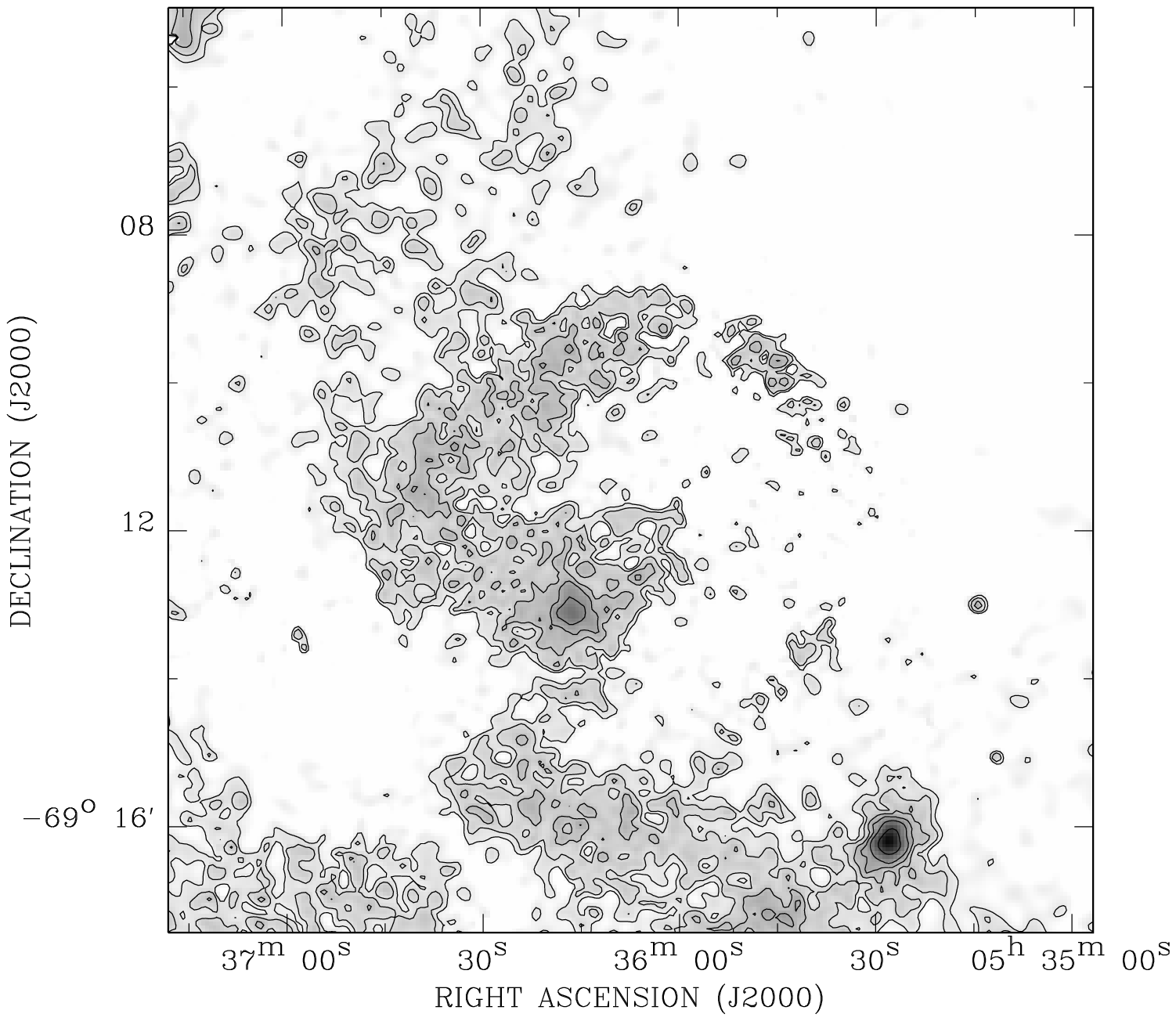,height=2.2in,angle=90,clip=}
\psfig{figure=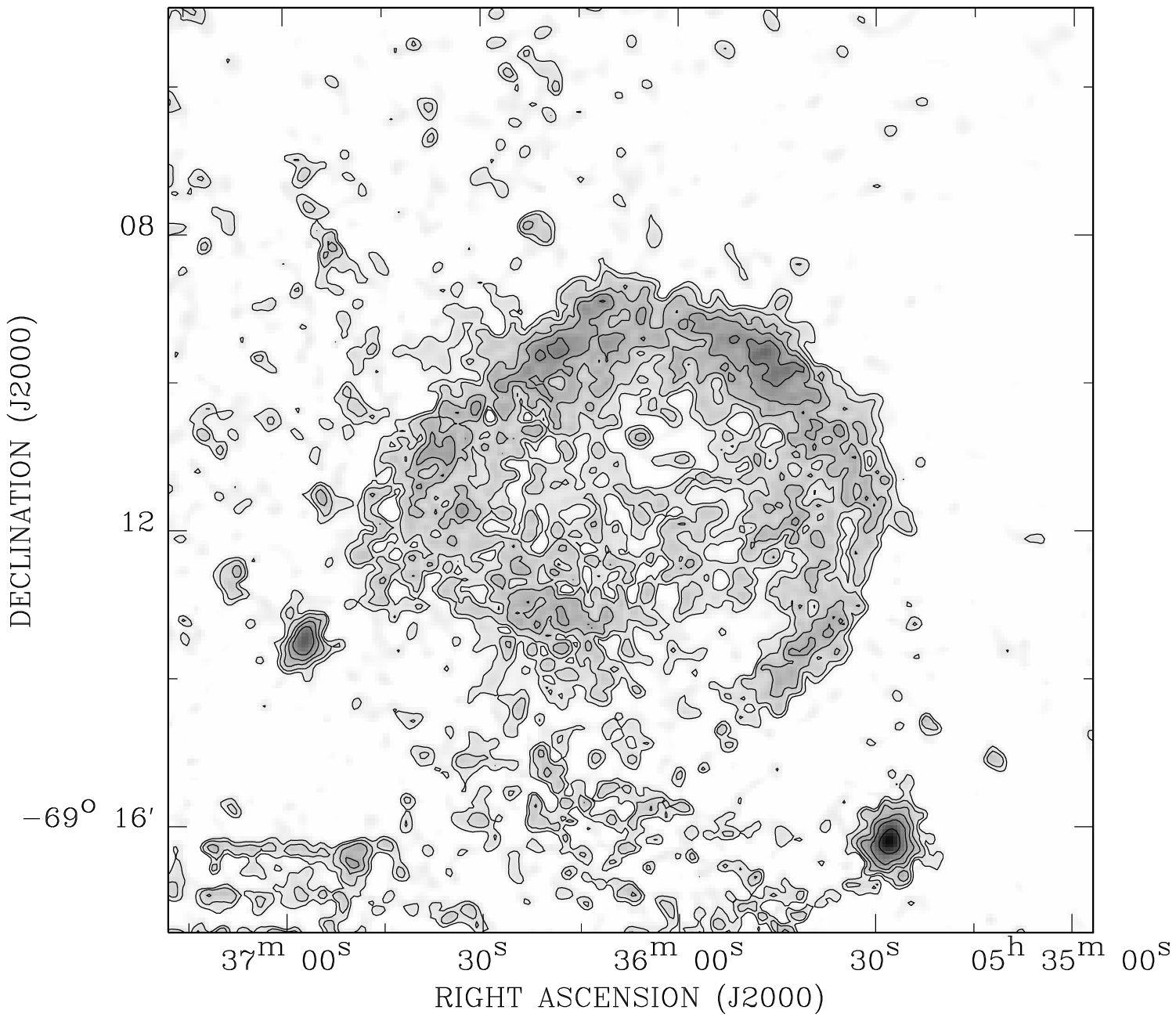,height=2.2in,angle=90,clip=}
\psfig{figure=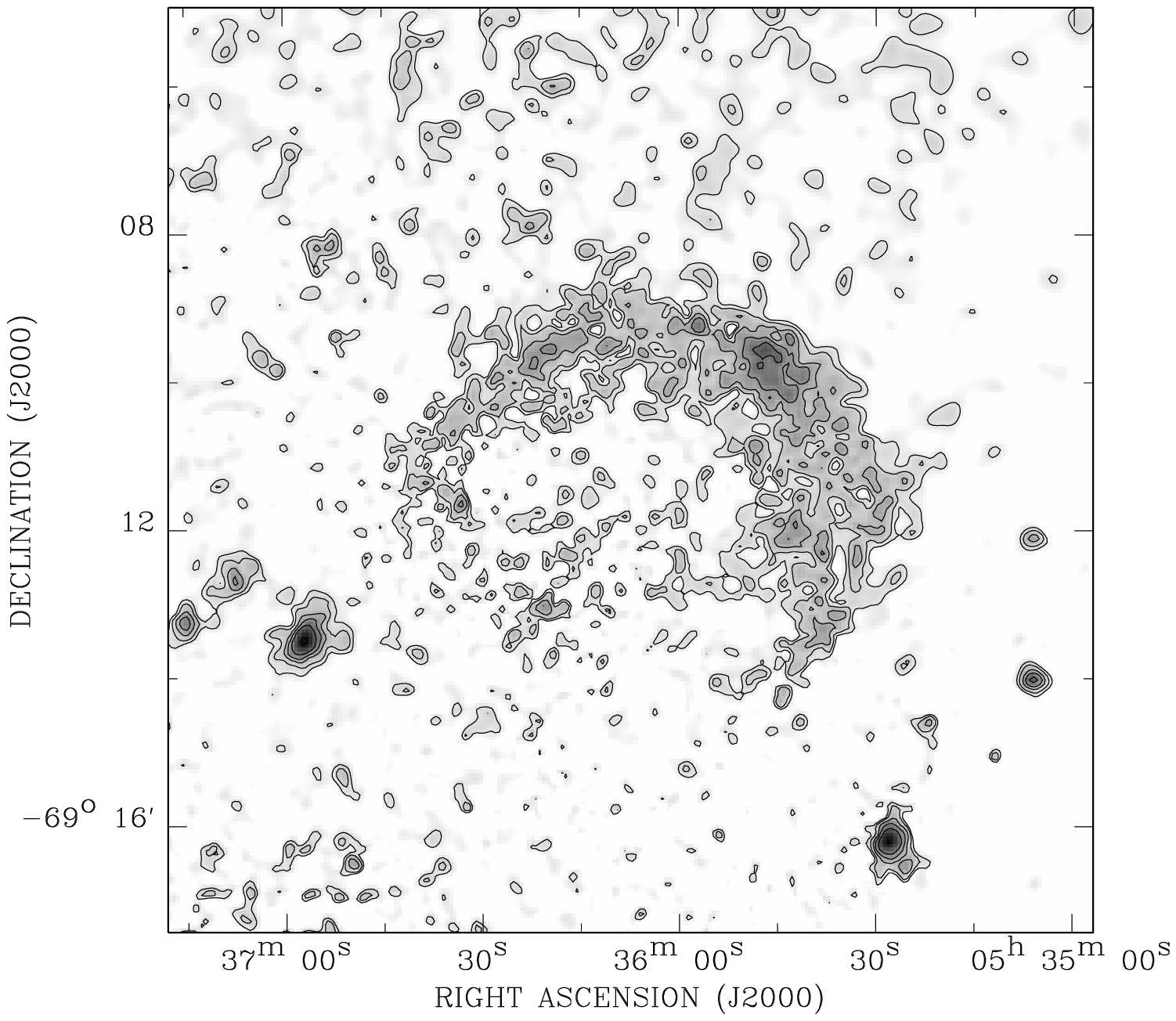,height=2.2in,angle=90,clip=}
}
\caption{Adaptively smoothed \emph{XMM-Newton} images of the central
region of 30 Dor C field in the (\emph{a}) 0.3--1, (\emph{b}) 1--2,
and (\emph{c}) 2--8~keV bands.  The X-ray images have been adaptively
smoothed using the same procedure as in Fig.~\ref{fig2}. The shading
is proportional to the logarithm of intensity and ranges from $3.2
\times 10^{-3}$ (white) to 0.48 (black) counts s$^{-1}$ arcmin$^{-2}$
in the 0.3--1 keV, from $2.7 \times 10^{-3}$ (white) to 0.41 (black)
counts s$^{-1}$ arcmin$^{-2}$ in the 1--2~keV, and from $3.1 \times
10^{-3}$ (white) to 0.14 (black) counts s$^{-1}$ arcmin$^{-2}$ in the
2--8~keV band images. Contours are drawn at (4.8, 6.4, 10, 16, 30, and
110) $\times 10^{-3}$ counts s$^{-1}$ arcmin$^{-2}$ in the 0.3--1~keV,
at (4.1, 5.4, 8.6, 14, 25, and 93) $\times 10^{-3}$ counts s$^{-1}$
arcmin$^{-2}$ in the 1--2~keV, and at (4.6, 6.1, 9.7, 15, 28, and 100)
$\times 10^{-3}$ counts s$^{-1}$ arcmin$^{-2}$ in the 2--8~keV band
images.}
\label{fig3}
\end{figure}

We also show adaptively smoothed images of the central
$12.\!^{\prime}5 \times 12.\!^{\prime}5$ region of the 30 Dor C field
in the soft ($0.3$--$1$~keV), medium ($1$--$2$~keV), and hard
($2$--$8$~keV) bands (Fig.~\ref{fig3}).  In these images, there is
evidence for spectral variations in the X-ray emission across the face
of the bubble.  While the western part of the bubble is easily seen in
the medium and hard band images, it is almost invisible in the soft
band image, clearly due to the X-ray absorption by foreground cool gas
(see also Dennerl et al. 2001 and Dunne et al. 2001).  There is a
region of localized CO emission projected onto the western part of 30
Dor C \citep{joh98}.  Assuming a $^{12}$CO(1-0) intensity of $9.1$ K
km s$^{-1}$ for the cloud (see Table~3 of Johansson et al. 1998) and a
CO to $\rm H_{2}$ conversion factor of $1.3 \times 10^{21} \, \rm mol
\, \rm cm^{-2} \, (\rm K \, \rm km \, \rm s^{-1})^{-1}$ for the LMC
\citep{isr97}, we estimate an $\rm H_{2}$ column density of $N_{\rm
H_{2}}\simeq 1.2 \times 10^{22} \, \rm mol \, cm^{-2}$ towards the
western part of 30 Dor C.  This absorption column density is
sufficient to absorb the soft X-ray emission from the western part of
bubble.  Intriguingly, in the hard band image, the western part of the
bubble appears much brighter than the eastern part, and is presumably
more luminous.

Finally, we note that J0536.9-6913 is almost invisible in the soft
band image, due to either absorption by gas in the LMC or absorption
intrinsic to the source.  This source also went undetected in the PSPC
observations reported by \citet{dun01}, which would require either an
increase in the absorption column density towards the source or a
decrease in the intensity of the source.

\subsection{Implementation of the Wind-Blown Bubble Spectral Model}
\label{sec-implementation}

In earlier studies of 30 Dor C with \emph{Einstein} and \emph{ROSAT},
the X-ray luminosity expected from the stellar wind-blown bubble model
of \citet{wea77} was often compared with that derived from spectral
fits to the data using a single temperature thermal plasma model
(e.g., Dunne et al. 2001).  The main disadvantage of this approach is
that the data are not directly compared to the model and so any test
of the model will depend on the measurements used to derive the
expected X-ray luminosity of the bubble (e.g., the number density and
expansion velocity of the H {\sc ii} region).  Also, the wide range of
gas temperatures inside the bubble gives rise to soft X-ray emission
in excess of that expected from a single temperature plasma model and
emission lines from a wider range of ionization states than those
expected from a single temperature thermal plasma model, which will
naturally result in an underestimate of the heavy element abundances.
The shortfall in the soft X-ray emission below $\sim 0.7$~keV from the
single temperature thermal plasma model can also lead to an
underestimate of the absorption column density, and hence the soft
X-ray luminosity.  In a more sophisticated analysis of 30 Dor C with
the \emph{Einstein} data, \citet{wh91} compared the IPC spectrum with
that expected from integrating the model over the bubble interior and
folding the result through the IPC response.  However, with the
limited quality of the IPC data, they were unable to constrain tightly
the basic spectral parameters such as temperature and absorption
column density, let alone to test other interesting parameters such as
the efficiency of thermal conduction and metal abundance of the X-ray
emitting gas.  In a more recent analysis of the \emph{Chandra} and
\emph{XMM-Newton} data, \citet{bam03,bam04} discovered nonthermal
X-ray emission from 30 Dor C, although the discovery was made using a
single temperature, non-equilibrium, ionized plasma model to describe
the thermal emission from the bubble.

Here we have compared the X-ray spectra of 30 Dor C directly with the
stellar wind-blown bubble model of Weaver et al. (1977; see also Mac
Low \& McCray 1988) by implementing it into the spectral analysis
package XSPEC.  The X-ray emission in their model arises primarily
from a region of thermally evaporated material (see Fig.~3 of Weaver
et al. 1977).  The temperature and density of the hot gas as a
function of radius $r$ within this region can be approximated as $T(r)
= T_{\rm c} \, (1-r/R)^{2/5}$ and $n(r) = n_{\rm c} \,
(1-r/R)^{-2/5}$, respectively, where $T_{\rm c}$ and $n_{\rm c}$ are
the corresponding central values, and $R$ is the outer radius of the
bubble.  The X-ray flux, $F_{\rm X}$, expected from the bubble is
\begin{eqnarray}
F_{\rm X} = \frac{1}{4 \pi D^{2}} \int \Lambda_{\rm X}[T(r)] \,
n(r)^{2} \, 4\pi r^{2} dr
\label{eqn-dem}
\end{eqnarray} 
where $\Lambda_{\rm X}(T)$ is the mean emissivity of the hot gas in
the X-rays and $D$ is the distance to the source.

We have calculated the X-ray spectrum with a multi-temperature,
variable heavy element abundance MEKAL (Mewe, Kaastra, \& Liedahl
1995) model, as implemented in XSPEC, using
\begin{eqnarray}
{\rm d}EM(T) = \frac{n^{2} \, 4\pi r^{2} dr}{4 \pi D^{2}} = K \, [ 2
(T/T_{\rm c})^{2} - (T/T_{\rm c})^{-1/2} - (T/T_{\rm c})^{9/2} ] \,
{\rm d}T/T_{\rm c}
\end{eqnarray}
for the differential emission measure ${\rm d}EM(T)/{\rm d}T$, which
is a function of temperature.  $K \, [= 10 \pi R^{3} n_{\rm c}^{2} /
(4 \pi D^{2}) = 15 V n_{\rm c}^{2} / (8 \pi D^{2})]$ is the
normalization of the spectrum in XSPEC.  In this form, the model
spectrum can be applied approximately to a slightly aspherical bubble
of an equivalent volume $V$.  

\subsection{Spectral Analysis}
\label{sec-spectral}

The MOS1 and MOS2 spectra of the diffuse emission in the eastern and
western parts of 30 Dor C (a total of 8 spectra) were simultaneously
fitted to the above wind-blown bubble model and a foreground
absorption of a column density of $N_{\rm H}$.  The absorption
cross-sections and atomic abundances were taken from \citet{mm83} and
\citet{ag89}, respectively.  While the foreground absorption was
allowed to be different between the eastern and western parts of 30
Dor C, we constrained the bubble model to have the same temperature,
metal abundance, and intensity in the eastern and western parts of 30
Dor C.  This model, which has 5 free parameters, gives a poor,
simultaneous fit to the MOS1 and MOS2 data with $\chi^{2} = 842.8$
(349 dof).  We also did not obtain a significant improvement in the
fit when we allowed the normalization of the bubble model to be
different between the MOS1 and MOS2 datasets.

An inspection of the spectra indicates that much of the deviation from
the best-fit bubble model is due to an excess of hard X-rays in the
spectra of the diffuse emission, particularly in the western part of
30 Dor C.  Such an hard X-ray excess can naturally be explained as
nonthermal X-ray emission, which can typically be described by a
power-law.  We therefore added a power-law component of photon index
$\Gamma$ to our model fits.  The normalization of the power-law
continuum was allowed to vary between the eastern and western parts of
30 Dor C, to account for the spatial variation in the hard X-ray
emission (Fig~\ref{fig3}).  In our initial modeling of the data, the
metal abundance was poorly constrained, with a lower limit of $\sim
0.1$ solar and an upper limit of several times solar.  We therefore
fixed the metal abundance at the known mean value of $0.5$ solar for
the ISM in the LMC \citep{rd92}.  In comparison to the bubble model,
this model (model B+P) gives a much improved, although still
statistically unacceptable, simultaneous fit to the MOS1 and MOS2 data
with $\chi^{2} = 459.8$ ($347$ dof; Table~\ref{tbl-1}, column 2).  The
absorbed $0.5$--$10$~keV flux in the western and eastern parts of 30
Dor C are $\sim 1.2 \times 10^{-12}$ and $\sim 1.3 \times 10^{-12}$
erg s$^{-1}$ cm$^{-2}$, respectively.  These fluxes are slightly
larger than the fluxes reported in \citet{bam03} for their regions
1--4, due to the differences in the sizes of the regions used for
extracting spectra between the present work and \citet{bam03}.  We
note that a marginal (at greater than $\sim 98$\% confidence --- a
$2\sigma$ result) improvement in the fit is obtained when we allowed
the normalization of the bubble model to be different between the
eastern and western parts of 30 Dor C.

The spectra of the diffuse X-ray emission in wind-blown bubbles have
often been fitted to single temperature plasma models in collisional
ionization equilibrium (e.g., Dunne et al. 2001), although with little
physical justification.  In order to compare with these earlier
observations and analyses, we have replaced the bubble model with a
single temperature MEKAL model.  We have also fixed the metal
abundance of the MEKAL at $0.5$ solar, since it is poorly constrained
by the data.  This model (model M+P) gives a similar quality fit to
the data as the B+P model, with $\chi^{2} = 463.7$ ($347$ dof;
Table~\ref{tbl-1}, column 3).  The best-fit equivalent hydrogen column
densities and normalizations of the power-law continua are comparable
to the values obtained from the fitting the data to the B+P model, and
the best-fit temperature of the MEKAL is consistent with the value
reported in \citet{bam03} for the thermal emission in their region 1.

We have also considered the possibility that the hard X-ray excess in
30 Dor C is attributable to a second thermal component that might be
caused by a recent supernova explosion, for example.  We replaced the
power-law component in the above M+P model with a second MEKAL model,
with the metal abundances of the two MEKALs fixed at the same value of
$0.5$ solar and the normalization of the high-temperature MEKAL
allowed to vary between the eastern and western parts of 30 Dor C.  We
find that this model provides a much worse fit to the data than do
either the B+P or M+P models, with $\chi^{2} = 550.7$ ($347$ dof).
This is due to the data at high energies being much too flat compared
to the model.  Although a significant (at greater than $99$\%
confidence) improvement in the fit is obtained when we allowed the
metal abundances of the two MEKALs to vary together, this model, with
one extra free parameter, still provides a worse fit to the data than
do either the B+P or M+P models, with $\chi^{2} = 509.2$ ($346$ dof).
Moreover, the metal abundance of $\sim 0.04$ solar is uncomfortably
small for the ISM in the LMC.  More sophisticated models, such as the
non-equilibrium, ionized (NEI) plasma codes as implemented in XSPEC
(see e.g., Borkowski, Lyerly, \& Reynolds 2001 for a detailed critique
of these models), also gave poorer fits to the data than do either the
B+P or M+P models.  We therefore conclude that the hard X-ray excess
in 30 Dor C is best described by a nonthermal emission component.

\begin{figure}
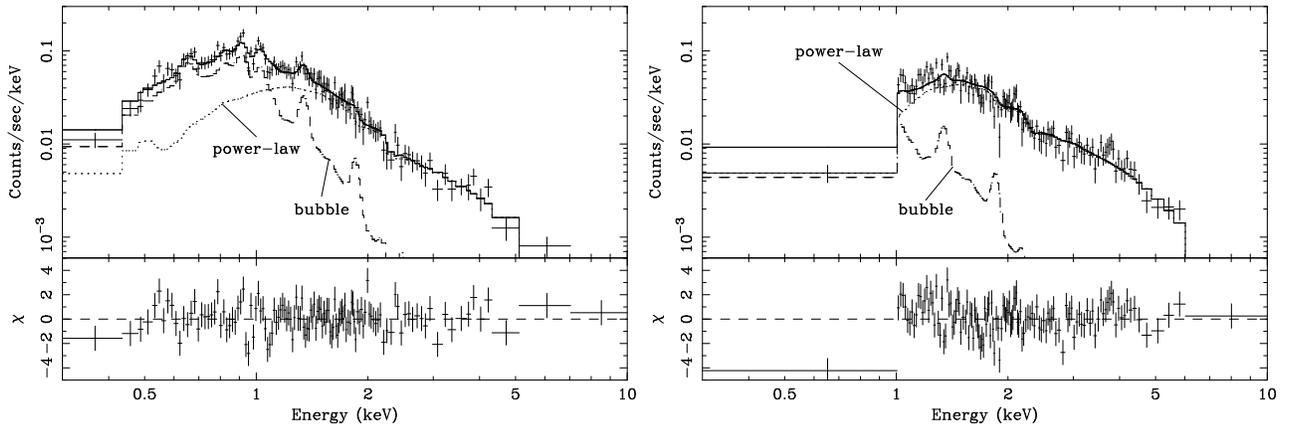
 
\centerline{ 
\psfig{figure=f4a.eps,height=2.2in,angle=-90,clip=}
\psfig{figure=f4b.eps,height=2.2in,angle=-90,clip=}
}
\caption{The co-added MOS spectra of the 30 Dor C superbubble.  The
upper panel shows the data together with the folded model (\emph{solid
lines}) comprising the best-fit wind-blown bubble spectrum
(\emph{dashed lines}) and power-law continuum (\emph{dotted lines}).
The residuals, in units of $\sigma$, from the best-fit model are shown
in the lower panel.  The data have been binned so that the
signal-to-noise ratio in each bin exceeds $4$. (\emph{a}) Eastern part
of superbubble. (\emph{b}) Western part of superbubble.}
\label{fig4}
\end{figure}

Because 30 Dor C is a massive star forming region with possible recent
core-collapse supernovae occurring inside the bubble, we have further
examined the data for an over-abundance of the $\alpha$-process
elements compared to Fe-like elements.  We did this by dividing the
elements into two groups: the $\alpha$-process elements O, Ne, Mg, Si,
S, Ar, and Ca; and the remaining Fe-like elements C, N, Na, Al, Fe,
and Ni.  Elements in each group were constrained to have the same
abundance relative to solar.  The abundances of He and the Fe-like
elements were fixed at the solar and $0.5$ solar values, respectively,
which are appropriate for the ISM in the LMC \citep{rd92}.  This model
(model vB+P) further improves the fit to the MOS1 and MOS2 data with
$\chi^{2} = 430.6$ ($346$ dof; Table~\ref{tbl-1}, column 3).  For
clarity, we show the co-added MOS spectra from the eastern and western
parts of the bubble, together with the residuals to this model fit
(Fig.~\ref{fig4}).  There are some large residuals around the energies
of the Al K$\alpha$ fluorescence line that arises from the detector
housing and the Au M-edge that arises from X-ray absorption in the
telescope mirror surface.  With the possible exception of Si, the
$\alpha$-process elemental abundance of $(1.6^{+0.6}_{-0.5})$ solar
(errors here and elsewhere in this paper are 90\% confidence for one
interesting parameter, $\Delta \chi^{2} = 2.706$) is higher than that
of the ISM in the LMC \citep{rd92}.  There is no improvement (at
greater than 90\% confidence) in the fit when we allow the abundance
of the Fe-like elements to vary.

Although the model spectrum can still be rejected statistically at
$\sim 99$\% confidence, we consider the fit to be reasonably good.
The model is certainly still an over-simplification.  For example, we
have assumed that the normalization of the soft bubble component is
the same for both the eastern and western parts of 30 Dor C.  This
assumption is problematic, however, since the power-law components of
the two parts are very different.  Because of the heavy absorption,
the data below $1$~keV does not allow us to independently constrain
the bubble component in the western part of 30 Dor C.  Dividing the
data into the two parts (east and west) is an oversimplification; in
general, the transition from east to west must be gradual and so the
parameters inferred from our model should be considered as averaged
values within the divided regions.  We have also not included any
systematic error in the data due to uncertainties in the calibration,
background subtraction, and modeling, which are difficult to quantify
at present.  We therefore decided not to add more complications (such
as varying abundances of individual elements) into the model.

\section{PHYSICAL PROPERTIES}
\label{sec-props}
The morphology of the diffuse X-ray emission in 30 Dor C is consistent
with that expected from the evaporation of cool gas into the hot
interior of a bubble created in the ISM by the winds of massive stars.
The close confinement of the limb-brightened X-ray emission by the
outer H$\alpha$-emitting shell indicates that the bubble is roughly
spherical.  Otherwise, the bubble has to be elongated almost exactly
along the line of sight.  We have therefore interpreted the X-ray
emission in terms of the stellar wind-blown bubble model of
\citet{wea77}.  

Assuming that the above wind-blown bubble model provides a reasonable
characterization of 30 Dor C, the physical properties of the enclosed
hot gas can be inferred from the spectral parameters obtained above.
We have adopted a notation similar to \citet{ss95}, where $t_{6} =
t/(10^{6} \, \rm yr)$ is the age of the superbubble in Myrs, $n_{0}$
is the ambient density of the ISM in cm$^{-3}$, $L_{38} = L_{\rm
mech}/(10^{38} \, \rm erg \, s^{-1})$ is the equivalent mechanical
power injected into the superbubble in units of $10^{38}$ erg
s$^{-1}$, and $\kappa_{0}$ is the ratio of the coefficient of thermal
conduction to the classical Spitzer value ($\kappa_{\rm sp} = 6 \times
10^{-7} \, T^{5/2}$ erg s$^{-1}$ K$^{-1}$ cm$^{-1}$).  The respective
central temperature and density of the hot gas are $T_{\rm c} = (5.3
\times 10^{6} \, {\rm K}) \{L_{38}^{8} n_{0}^{2} / (t_{6}^{6}
\kappa_{0}^{10})\}^{1/35}$ and $n_{\rm c} = (1.6 \times 10^{-2} \,
{\rm cm}^{-3}) \{L_{38}^{6} n_{0}^{19} \kappa_{0}^{10} /
t_{6}^{22}\}^{1/35}$, assuming a superbubble radius $R = (66 \, {\rm
pc}) \{L_{38} t_{6}^{3} / n_{0}\}^{1/5}$ (Shull \& Saken 1995; see
also Mac Low \& McCray 1988).  We can rearrange the above equations
for $T_{\rm c}$, $n_{\rm c}$, and $R$ to give
\begin{eqnarray}
n_{0} = \frac{(n_{\rm c}/1.6 \times 10^{-2} \, {\rm cm}^{-3}) \,
(T_{\rm c}/5.3 \times 10^{6} \, \rm K)}{(R/66 \, {\rm pc})^{2}} \,
t_{6}^{2}, \label{eqn-n0}
\end{eqnarray}
\begin{eqnarray}
L_{38} = \frac{(n_{\rm c}/1.6 \times 10^{-2} \, {\rm cm}^{-3}) \,
(R/66 \, {\rm pc})^{3} \, (T_{\rm c}/5.3 \times 10^{6} \, \rm
K)}{t_{6}}, \label{eqn-mechlum}
\end{eqnarray}
and
\begin{eqnarray}
\kappa_{0} = \frac{(n_{\rm c}/1.6 \times 10^{-2} \, {\rm cm}^{-3}) \,
(R/66 \, {\rm pc})^{2}}{(T_{\rm c}/5.3 \times 10^{6} \, \rm K)^{5/2}
\, t_{6}}. \label{eqn-kappa0}
\end{eqnarray}

The parameters on the right side of
equations~(\ref{eqn-n0})--(\ref{eqn-kappa0}) can be inferred from
either the spectral fits with the bubble model ($n_{\rm c}$ and
$T_{\rm c}$) or other independent measurements ($R$ and $t_{6}$).  The
central temperature and normalization of the best-fit bubble model
(model vB+P) are $T_{\rm c} = 7.4 \times 10^{6} \, \rm K$ and $K = 6.1
\times 10^{10} \, \rm cm^{-5}$, respectively.  This value of $K$ gives
a central density of $n_{\rm c} = 1.4 \times 10^{-2} \, \rm cm^{-3}$
and a mean emission measure of $\simeq 0.08$ cm$^{-6}$ pc, assuming
the bubble to be a prolate spheroid of dimension $50 \, \rm pc \times
45 \, \rm pc$.  Adopting a mean radius of $47 \, \rm pc$ for the
bubble, which is the radius of the sphere that has the same volume as
the prolate spheroidal bubble, gives an ambient density of $n_{0} =
(2.4 \, \rm cm^{-3}) \, t_{6}^{2}$, a mechanical luminosity of $L_{\rm
mech} = (4.3 \times 10^{37} \, \rm erg \, \rm s^{-1}) / t_{6}$, and a
coefficient of thermal conduction of $\kappa_{0} \simeq 0.19 / t_{6}$
of the Spitzer value.  The mass of gas that has been swept up is then
$M_{\rm S} = \mu n_{0} V \simeq (3.3 \times 10^{4} \, M_{\odot}) \,
t_{6}^{2}$, where $\mu = (14/11) \, m_{\rm H}$ is the mass per
particle, $m_{\rm H}$ is the mass of a hydrogen atom, and $M_{\odot}$
is the mass of the Sun.  Following \citet{ss95}, we also estimate the
mass of X-ray emitting gas at temperatures $\simgreat 10^{5}$~K as
$M_{\rm X} \simeq (375/156) \mu n_{\rm c} V \simeq 420 \, M_{\odot}$.
Finally, the thermal and kinetic energy of the superbubble are $E_{\rm
TH} \simeq (5/11) \, L_{\rm mech} t \simeq 6.2 \times 10^{50} \, \rm
erg$ and $E_{\rm KE} \simeq (15/77) \, L_{\rm mech} t \simeq 2.7
\times 10^{50} \, \rm erg$ respectively, and the thermal pressure of
the hot gas is $P_{\rm TH} \simeq (2/3) \, (E_{\rm TH}/V) \simeq 3.3
\times 10^{-11} \, \rm dyne \, cm^{-2}$.  From the total energy, we
infer that one or two core-collapse supernovae could have occurred
within the superbubble, since each supernova supplies roughly
$10^{51}$ erg (e.g., Woosley \& Weaver 1986).  W-R stars are the most
probable core-collapse supernovae progenitors in LH~90, and we may
expect these stars to have a mass distribution similar to that of the
W-R stars in the Galaxy, i.e., masses in the range from $\sim 5$ to
$\sim 30 \, M_{\odot}$ (e.g., Cherepashchuk 1991). Stars with masses
in the range from $\sim 12$ to $\sim 30 \, M_{\odot}$ are expected to
undergo nuclear fusion of the $\alpha$-elements and yield between
$\sim 0.15$ and $\sim 4 \, M_{\odot}$ of oxygen \citep{ww95,thi96}.
From the best-fit $\alpha$-process elemental abundance of $1.6$ solar,
we infer an oxygen/hydrogen mass ratio of $0.022$ for the X-ray
emitting gas in the bubble.  The ISM in the LMC has an
$\alpha$-process elemental abundance of $\sim 0.5$ solar or an
oxygen/hydrogen mass ratio of $6.7 \times 10^{-3}$.  Thus, the bubble
must have been enriched by $\sim 5 \, M_{\odot}$ of oxygen, assuming
that $\simeq 75$ percent of the mass of the X-ray emitting gas is
hydrogen.  At least 2--3 high ($\simgreat 20 \, M_{\odot}$) mass,
core-collapse supernovae are needed to explain the oxygen abundance in
the bubble, and the total energy supplied exceeds the value inferred
from the spectral fits.

\section{CHECKING THE PHYSICAL PARAMETERS WITH MULTIWAVELENGTH DATA}
\label{sec-multiwavelength}

To make our results a bit more quantitative, we need to know the age
of the bubble, which is still uncertain.  Following \citet{chu95a}, we
use the observed expansion velocity of $V_{\rm exp} = (0.59 \, \rm km
\, s^{-1}) \, R_{\rm pc} / t_{6} \sim 45$ km s$^{-1}$ \citep{dun01},
where $R_{\rm pc} \simeq 47$ in units of pc, to derive a bubble age of
$t \sim 6.2 \times 10^{5}$ years.  This value of $t$ is probably a
lower limit to the true age of the bubble since the expansion is only
detected in parts of the bubble (Y.-H. Chu 2004, private
communication).  The energy input from the OB association also tends
to increase with time \citep{ss95}.  The ages of the most massive
stars in the OB association give an age of $\sim 3$--$4$ Myr
\citep{tes93}, although there are subclusters that are older (up to
$\sim 7$~Myr). This kind of progressive star formation is also seen in
the central OB association of the nearby 30 Dor nebula, typically with
increasing strength; i.e., the most massive stars are formed later
(e.g., Selman et al. 1999). Therefore, we adopt a mean age of the OB
association as $\sim 4$ Myrs. The inferred ambient density of $n_{0} =
38 \, \rm cm^{-3} $ is reasonable for the expected high density ISM
near 30 Dor C.

The mean emission measure of the H$\alpha$-emitting shell is $\sim
2000$ cm$^{-6}$ pc \citep{dun01}.  Adopting a minimum path length of
$20 \, \rm pc$ through the H$\alpha$ emission region, which is twice
as large as the width of the optical line-emitting filaments in
Figure~\ref{fig2} (see also Fig.~1\emph{l} of Dunne et al. 2001),
gives an upper limit to the mean electron density of $n_{\rm e}
\simless 10 \, \xi_{\rm H\alpha}^{-1/2} \, \rm cm^{-3}$ for the
ionized shell, where $\xi_{\rm H\alpha}$ is the filling factor of the
H$\alpha$-emitting gas.  The thermal pressure of the ionized shell is
approximately $2 n_{\rm e} k T_{\rm e} \simless 2.8 \times 10^{-11} \,
\xi_{\rm H\alpha}^{-1/2} \, \rm dyne \, \rm cm^{-2}$, where $T_{\rm e}
\simeq 10^{4}$ K and $k$ is the Boltzmann constant.  This value is
within a factor of three of the thermal pressure derived from the
X-ray data for filling factors in the range $0.1$--$1$.

The mechanical luminosity required to power the bubble is $L_{\rm
mech} = 1.1 \times 10^{37} \, \rm erg \, \rm s^{-1}$.  There are $26$
spectroscopically indentified O-type stars and $7$ confirmed W-R stars
in LH~90 \citep{tes93}.  While the mass loss rates of W-R stars can be
an order of magnitude higher than those of O-type stars (e.g., Nugis
\& Lamers 2000), the lifetime of these stars are fractions of a
million years, i.e., much shorter than the probable age of the bubble,
and so their energy input to the bubble is negligible compared with
that from the O-type stars.  Adopting a mean mass loss rate of
$\dot{M} \sim 10^{-6} \, M_{\odot} \, \rm yr^{-1}$ (de Jager,
Nieuwenhuijzen, \& van der Hucht 1988) and terminal velocities in the
range from $V_{\infty} \sim 1000$ to $3000 \, \rm km
\, \rm s^{-1}$ (Prinja, Barlow, \& Howarth 1990), which are typical
values for the O-type stars in LH~90, we calculate an integrated wind
luminosity of $L_{\rm w} = \Sigma (1/2) \dot{M} V_{\infty}^{2} \sim
1$--$7 \times 10^{37} \, \rm erg \, \rm s^{-1}$.  Therefore, the
O-type stars in the stellar clusters of LH~90 could easily supply the
mechanical luminosity required to power the bubble, without the
contributions from a few supernovae.  If the bubble is much older than
$4$~Myr, then the mechanical energy input to the bubble from the
O-type stars could greatly exceed the value of $L_{\rm mech}$ inferred
from the spectral fits.  The best-fit bubble parameters give an
estimate of the total energy input of $L_{\rm mech} t \simeq 1.4
\times 10^{51} \, \rm erg$.  This energy input is inferred from the
density and temperature structure of the bubble, and not from its
mechanical luminosity and age.

There is clearly discrepancy between the expected mechanical energy
input from the OB association and that inferred from the bubble
model. This apparent oversupply of mechanical energy is a common
problem in the existing studies of several other structures around OB
associations (e.g., Oey 1996; Naz\'{e} et al. 2001; Cooper et
al. 2003).  In all these studies, including the present one, a
potentially very important energy mechanism --- dust grain cooling ---
is not considered. One expects dust grains to be mixed with the
X-ray-emitting gas due to the thermal evaporation and supernova
ejecta. With a normal interstellar dust grain population, the cooling
rate of dust grains heated by collisions with hot electrons can be
substantially greater than the radiative cooling of the hot gas itself
(e.g., Dwek \& Arendt 1992 and references therein). Of course, dust
grains subject to the sputtering. Their survival timescale depend on
both the grain sizes and the ambient hot gas density. With the large
uncertainties in these parameters, we find that it is conceivable for
dust grains to be the major coolant of a superbubble.

We find a coefficient of thermal conduction of $\kappa_{0} = 0.05 \pm
0.04$ of the classical Spitzer value.  Thermal conduction can be
reduced by the presence of magnetic fields (e.g., Chandran \& Cowley
1998): the thermal conductivity can approach $\sim 20$\% of the
Spitzer value due to the chaotic transverse wandering of the magnetic
field lines, although this value could be uncertain by a factor of two
\citep{nm01}.  The theoretical value of $\kappa_{0}$ is consistent
with our measurement, given the uncertainties in the bubble age.
Thermal conduction is unlikely to be saturated (e.g., Cowie \& McKee
1977), since the mean free path of a charge particle is much smaller
than the radius of the bubble.

As expected, the best-fit equivalent hydrogen column densities toward
the eastern and western parts of 30 Dor C are significantly higher
than the equivalent Galactic H~{\sc i} column density toward 30 Dor C
of $N_{\rm H} (\rm Gal) = 6.4 \times 10^{20}$ cm$^{-2}$ \citep{dl90}.
We therefore use the optical extinction of stars in the OB association
to independently estimate the total (atomic plus molecular) column
density toward 30 Dor C.  The mean interstellar extinction for the
association as a whole is $E_{\rm B-V} \sim 0.4 \, \rm mag$ and
reaches a maximum of $E_{\rm B-V} \sim 0.8 \, \rm mag$ in the western
part \citep{tes93}.  Adopting a value of $E_{\rm B-V} = 0.08 \, \rm
mag$ for the Galactic foreground extinction toward 30 Dor C
\citep{xu92} and assuming gas-to-dust ratios of $N_{\rm H}/E_{\rm B-V}
= 4.8 \times 10^{21} \, \rm cm^{-2} \, \rm mag^{-1}$ (Bohlin, Savage,
\& Drake 1978) and $N_{\rm H}/E_{\rm B-V} = 2.4 \times 10^{22} \, \rm
cm^{-2} \, \rm mag^{-1}$ \citep{fit85} for the Galaxy and LMC
respectively, we estimate a mean column density of $N_{\rm H} \sim 8.1
\times 10^{21} \, \rm cm^{-2}$ toward the OB association and a maximum
column density of $N_{\rm H} \sim 1.8 \times 10^{22} \, \rm cm^{-2}$
toward the western part.  While these values appear to be larger than
our measurements, we had assumed solar abundances for the atomic
absorption cross-sections in the spectral fits.  Almost all of the
absorption cross-section at $\sim 1$ keV is provided by metals and so,
for a metal abundance of $0.5$ solar, the measured X-ray absorption
column densities towards the eastern and western parts of the bubble
are $N_{\rm H} \sim 10^{22}$ cm$^{-2}$ and $\sim 2 \times 10^{22}$
cm$^{-2}$, respectively.  These X-ray absorption column densities are
consistent with the absorption column densities estimated from the
optical extinction of stars in the OB association.  The difference in
the measured X-ray absorption column densities of $N_{\rm H} \sim
10^{22}$ cm$^{-2}$ is consistent with the H$_{2}$ column density
inferred from the CO emission projected onto the western part of 30
Dor C (see \S~\ref{sec-morphology}).

\section{NATURE OF THE NONTHERMAL X-RAY EMISSION}
\label{sec-nonthermal}

\subsection{Synchrotron Emission}
\label{sec-synchrotron}

For energy-conserving shocks, the H$\alpha$ expansion velocity of
$45$~km~s$^{-1}$ \citep{dun01} corresponds to a shock velocity of
$\sim 60$~km~s$^{-1}$ (e.g., Chu et al. 1995b), and gas passing
through the shock would only be heated to temperatures of $\simeq
10^{5}$~K (e.g., Hollenbach \& McKee 1979), which is well below the
temperature of the X-ray emitting gas.  The observed [S {\sc
ii}]/H$\alpha$ flux ratio of $0.3$ \citep{mat85} is also below that
expected from shocked gas (e.g., Long et al. 1990) and, contrary to
the observations, the bulk of the nonthermal X-ray emission is
expected to originate in the eastern part of 30 Dor C since the
non-thermal radio emission is much stronger on the eastern side of the
bubble than on the western side.  Thus, we conclude that any recent
supernova explosion must have occured deep inside the bubble,
producing little or no observable nonthermal X-ray synchrotron
radiation.  \citet{bam03,bam04} explained the the apparently
nonthermal component in the X-ray spectrum of 30 Dor C as synchrotron
radiation from relativistic electrons.  If this is true, the required
electron energies should be a few hundred TeV for a typical
interstellar magnetic field of order $B = 3 \times 10^{-6}$~G.
Although low (up to a few tens of km s$^{-1}$) velocity shoulders are
visible on the H$\alpha$ emission lines from various parts of the
bubble (Y.-H. Chu 2004, private communication), there is a complete
lack of emission with velocities greater than $100$ km s$^{-1}$ from
the recessional velocity of the LMC and so the shock velocities are
certainly too low to produce such energetic particles.

The most obvious site for an effective particle acceleration should be
the reverse shock, which separates the free-streaming stellar wind
from the region of thermally evaporated material ($R_{1}$ in Fig.~1 of
Weaver et al. 1977).  For a mean mass loss rate of $\dot{M} \sim
10^{-6} \, M_{\odot} \, \rm yr^{-1}$, a terminal velocity of
$V_{\infty} \sim 2000 \, \rm km \, \rm s^{-1}$, an ambient density of
$n_{0} = 38$ cm$^{-3}$, and a bubble age of $4$ Myrs, the inner shock
radius is roughly $7$~pc \citep{wea77}. The time scales for the
relativistic electrons to diffuse from the reverse shock to the edge
of the bubble would have to be shorter than their synchrotron
half-lives, which are of order $(5 \times 10^{3} \, \rm years) \, (B/3
\times 10^{-6} \, \rm G)^{-3/2}$ (e.g., Lang 1974, pp. 31-32).  The
relativistic electrons would therefore have to travel in the bubble at
velocities that are an order of magnitude faster than the sound
velocity of $\sim 10^{2}$--$10^{3}$ km s$^{-1}$ and two orders of
magnitude faster than the Alfv\'{e}n velocity of $B/(9.6 \pi \mu
n_{\rm c})^{1/2} \sim 30$ km s$^{-1}$, which is untenable (e.g., Jaffe
1977). It is possible that the electrons could be accelerated by
turbulent motions in the bubble itself.  The turbulent velocities
would be of order or greater than the Alfv\'{e}n velocity, since
otherwise the magnetic field would damp the turbulent motions (e.g.,
Jaffe 1977), and so the turbulent kinetic energy involved would be a
few $10^{48}$ erg.  The electrons must be accelerated on time scales
that are shorter than their radiative lifetimes and so the power
dissipated in this process is of order $10^{37}$ erg s$^{-1}$.  While
this value is comparable to the mechanical wind luminosity of the
early-type stars, the ``second-order'' acceleration by turbulence is
probably too inefficient.
 
\subsection{Inverse Compton Radiation}

The inverse Compton (IC) scattering of Cosmic Microwave Background
(CMB) and other ambient photons (e.g., infrared photons from
reprocessed starlight) by relativistic electrons will also generate
nonthermal X-ray emission.  The upscattering of such photons 
to $5$~keV requires electrons with energies in the range from
$\sim 0.1$ to $\sim 1$~GeV.  These electrons radiate synchrotron
emission at frequencies of MHz to tens of MHz in magnetic fields of
order $3 \times 10^{-6} \, \rm G$.  For CMB photons, the ratio of
inverse Compton-scattered X-ray flux to radio synchrotron flux is
\begin{eqnarray} 
\frac{S_{\rm x}}{S_{\rm r}} \simeq 5.1 \times 10^{-18} \,
B^{-(1+\alpha)} \, \frac{b(s)}{a(s)} \left( 13.6 \times 10^{4} \,
\frac{\nu_{\rm r}}{\nu_{x}}\right)^{\alpha} \label{eqn-ic}
\end{eqnarray}
(e.g., Blumenthal \& Gould 1970), where $S_{\rm x}$ and $S_{\rm r}$
are the flux densities at frequencies $\nu_{\rm x}$ and $\nu_{\rm r}$,
respectively.  The functions $b(s)$ and $a(s)$ are given in
\citet{bg70} for a power-law distribution of electron energies with
index $s$.  The radio synchrotron spectrum should have the same
spectral index as the IC emission and so we have assumed a spectral
index of $\alpha = (s - 1)/2 = 1.5$ for the radio synchrotron
spectrum.  The ratio of $b(s)$ to $a(s)$ is $\simeq 420$ for this
value of $\alpha$.  We have also assumed that the magnetic field
strength is $3 \times 10^{-6} \, \rm G$ and that the electron
distribution is an unbroken power-law extending from Lorentz factors
of few times $10^{3}$, which will upscatter CMB photons to X-ray
wavelengths, to Lorentz factors of order $10^{4}$, which will produce
$843$~MHz radio emission via synchrotron radiation.  This last
assumption may not be correct, however, since the nonthermal X-ray
emission is stongest in the western part of the bubble where there is
no evidence for nonthermal radio emission \citep{mat85}.  Nonetheless,
the total (thermal plus synchrotron) emission at $843$~MHz is $1.2$~Jy
\citep{mat85} and so the expected 1~keV to $843$~MHz flux density
ratio of $\sim 1.4 \times 10^{-6} \, (B/3 \times 10^{-6} \, \rm
G)^{-5/2}$ gives an upper limit to the IC emission at 1~keV of $\sim
2.5 \times 10^{-3} \, (B/3 \times 10^{-6} \, \rm G)^{-5/2}$ photons
cm$^{-2}$ s$^{-1}$ keV$^{-1}$.  Although this value is consistent with
the flux densities of the power-law emissions from the eastern and
western parts of 30 Dor C, the IC emission is probably much fainter
than this, since the flat ($\alpha \sim 0$) radio spectrum is an
indication that the radio emission is mostly thermal \citep{mat85}.

The upscattering of infrared photons to X-ray energies requires
electrons with Lorentz factors of a few hundred.  For a power-law
distribution of electron energies with index $s \simeq 4$, we would
expect the number of electrons capable of upscattering infrared
photons to X-ray energies to be $\sim 10^{2}$ times the number of
electrons capable of upscattering CMB photons to X-ray energies.  The
$42.5$--$122.5 \, \mu$m luminosity of 30 Dor C can be estimated from
\begin{eqnarray} 
L_{\rm fir} = 4 \pi D^{2} \, 1.26 \times 10^{-11} (2.58 \, f_{60} +
f_{100}) \, {\rm erg} \, {\rm s}^{-1}
\end{eqnarray}
(\emph{Cataloged Galaxies and Quasars Observed in the IRAS Survey},
Appendix B), where $f_{60}$ and $f_{100}$ are the \emph{IRAS} flux
densities in Jy at $60\mu$m and $100\mu$m, respectively, and $D$ is
the distance to the source in cm.  The extended infrared emission
measured by \emph{IRAS} is roughly $15$ and $30$ Jy at $60$ and $100
\mu$m, respectively \citep{lm91}, and so we estimate a $42.5$--$122.5
\, \mu$m luminosity of the order of a few $10^{38}$ erg s$^{-1}$.
Thus, the mean infrared energy density of the bubble is of the order
$10^{-13}$ erg cm$^{-3}$ and could be higher if the infrared emission
is concentrated in the outer shell.  If we adopt a mean photon energy
of $\epsilon = 0.018$ eV for the infrared emission, then the total
infrared photon number density is $\sim 3.5$ cm$^{-3}$.  The CMB
photons have, in contrast, a mean photon energy of $\epsilon = 6.25
\times 10^{-4}$ eV and a total photon number density of $400$
cm$^{-3}$ (e.g., Gaiser, Protheroe, \& Stanev 1998). From equation (6)
of \citet{gai98} and assuming that the electron energies are
distributed as a power-law of index $s \simeq 4$, we find that the IC
emission at a fixed photon energy scales roughly as $\epsilon^{3/2} \,
n_{\rm ph}$, where $n_{\rm ph}$ is the photon number density of the
radiation under consideration.  So we would expect the X-ray emission
from IC-scattered infrared photons to be $\sim 30$ percent higher than
that from the upscattering of the CMB photons.  However, the optical
radiation from stars should have an energy density greater than that
of the infrared emission and could be IC-scattered to X-ray energies
by electrons with Lorentz factors of $\sim 50$.  If we assume that the
mean energy of the optical photons is $\epsilon \simeq 3.1$ eV, then
the total photon number density of the optical radiation should be
$\simgreat 0.02$ cm$^{-3}$ and the X-ray emission from IC-scattered
optical photons should be an order of magnitude higher than that from
the upscattered infrared and CMB photons.  Unless the particle energy
distribution is far from the power law (e.g., a Maxwellian
distribution) and/or the particle population is greatly enhanced at
the shell, the nonthermal X-ray emission should then be concentrated
in the central region of 30 Dor C, where the optical emission is
greatest.  This is inconsistent with the observed limb-brightened hard
X-ray emission (Fig.~\ref{fig3}).  A strong shock, as expected inside
the bubble, will also produce electron spectra with indexes of $s \sim
2$ (e.g., Blandford \& Eichler 1987), which is totally at odds with
the spectral index of the nonthermal X-ray emission. On the other
hand, if the relativistic particles represent the thermalized wind
material from young pulsars during the lifetime of the bubble, then
the particle energies could be distributed as a Maxwellian of peak
energy $\gamma \sim 10^{2}$--$10^{3}$ \citep{at94,ato99}.  Therefore,
we cannot rule out the possibility that the bulk of the nonthermal
X-ray emission from the bubble is due to IC radiation.

\subsection{Nonthermal Bremsstrahlung Radiation}

Nonthermal particles generate bremsstrahlung in regions where the
energy losses due to ionization and Coulomb collisions are
significant.  The energy flux of the bremsstrahlung radiation is
$\simless 10^{-5}$ times the total energy input to the particles
(e.g., Dogiel et al. 2002), i.e., a large amount of energy is required
to maintain the particles against Coulomb/ionization losses to the
emitting region.  The best-fit bubble model (model vB+P) has power-law
components with unabsorbed $0.1$--$10$~keV luminosities of $1.2 \times
10^{36}$ and $2.8 \times 10^{36}$ erg s$^{-1}$ in the eastern and
western parts of the bubble, respectively.  The required energy input
is therefore $\simgreat 10^{41}$--$10^{42}$ erg s$^{-1}$.  Although
the necessary energy input can be less than this for a quasi-thermal
distribution of electrons formed in regions of in situ acceleration
\citep{dog02}, it is still orders of magnitude greater than the
integrated wind luminosity of $\sim 1$-$7 \times 10^{37}$ erg s$^{-1}$
from the O-type stars in LH~90.  A recent (within the last a few $\sim
10^{3}$ years; Chu \& MacLow 1990) supernova explosion could provide
the energy input to the bubble, but this would mean that we are seeing
the event by chance.  The 26 O-type stars in LH~90 have masses in the
range from $20$ to $40 \, M_{\odot}$ (e.g., Walborn et al. 1995,
1999). If we assume a Spitzer initial mass function for the stars in
LH~90, then we would expect $\sim 12$ O-type stars with masses in the
range from $40$ to $60 \, M_{\odot}$.  Although some of these stars
would undoubtedly have evolved into W-R stars, the remainder would
have exploded as supernovae (stars with masses greater than $\sim 60
\, M_{\odot}$ do not explode as supernovae, but instead collapse into
a black-hole).  Thus, there must have been at least $\sim 5$--$6$
supernovae in 30 Dor C, and so the probability of seeing a supernova
within the last $10^{4}$ years is approximately $1$ percent, assuming
that the age of the bubble is $4$ Myr.

\section{CONCLUSIONS}

In summary, the nature of the apparent nonthermal hard X-ray component
is still uncertain.  The expansion of the bubble is much too slow in
order to produce the energetic particles necessary for nonthermal
X-ray synchrotron emission.  Although such energetic particles can be
produced at the inner reverse shock, they cannot cross the bubble
within their synchrotron half-lives.  Nonthermal bremsstrahlung is an
inefficient process, requiring at least three orders of magnitude more
energy than can be supplied by the O-type stars in LH~90.  A recent
supernova could supply the necessary energy, but this would mean that
we are seeing the event by chance (a probability of $<1$ percent).
The IC scattering of CMB and infrared photons cannot account for all
of the nonthermal X-ray emission unless the radio flux is purely
synchrotron radiation, which seems unlikely.  While the X-ray emission
from IC-scattered optical photons should be much higher than that from
IC-scattered CMB and infrared photons, it should be concentrated in
the central region of 30 Dor C, where the optical photon density is
greatest, in sharp contrast to the data.  The nonthermal X-ray
emission could instead be due to the IC scattering of CMB and infrared
photons by relativistic particles originating in young pulsar wind
material with energies peaking between $\gamma \sim 10^{2}$--$10^{3}$.
High spatial resolution X-ray imaging, which can be afforded by an
on-axis \emph{Chandra} observation, will be particularly useful to
further the understanding of the component.  Such an observation will
allow for an accurate determination of the size and geometry of the
X-ray-emitting region and for a detailed comparison with images in
other wavelength bands.  A substantial fraction of the expected energy
input from stellar winds and supernovae may be radiated in the
infrared by dust grains entrained in the hot gas.

We thank R. Chris Smith for providing his optical image of 30 Dor C in
computer-readable format, Glenn Allen for discussions on the
nonthermal emission, and the referee for insightful comments on an
earlier draft of the paper.  This research is supported by the NASA
LTSA grant NAG5-8935.


\clearpage

\begin{deluxetable}{ccccc}
\footnotesize
\tablecolumns{5}
\tablecaption{Spectral Fits to the Integrated Emission from 30 Dor C
in the $0.3$--$10$ and $1$--$10$~keV bands. \label{tbl-1}}
\tablewidth{0pt}
\tablehead{
\multicolumn{2}{c}{} & \multicolumn{3}{c}{Model\tablenotemark{a}} \\
\cline{3-5} \\
\multicolumn{2}{c}{Parameter} &
\colhead{B+P} &
\colhead{M+P} &
\colhead{vB+P}}
\startdata

$N_{\rm H, east}$\tablenotemark{b} & ($10^{21}$ cm$^{-2}$) & $5.4^{+1.1}_{-0.6}$ & 
        $5.94^{+0.55}_{-0.65}$ & $4.08^{+0.59}_{-0.56}$ \\
$N_{\rm H, west}$\tablenotemark{c} & ($10^{21}$ cm$^{-2}$) & $11.41^{+0.96}_{-0.71}$ & 
        $11.90^{+0.66}_{-0.68}$ & $10.18^{+0.73}_{-0.69}$ \\
$kT$\tablenotemark{d} & (keV) & $0.359^{+0.046}_{-0.068}$ & 
        $0.186^{+0.014}_{-0.008}$ & $0.64^{+0.20}_{-0.12}$ \\
$Z$\tablenotemark{e} & ($Z_{\odot}$) & $0.50$\tablenotemark{f} & 
        $0.50$\tablenotemark{f} & $0.50$\tablenotemark{f} \\
$Z_{\alpha}$\tablenotemark{g} & ($Z_{\odot}$) & $0.30$\tablenotemark{f} & 
        $0.50$\tablenotemark{f} & $1.63^{+0.58}_{-0.45}$ \\
$K_{\rm thermal}$\tablenotemark{h} & ($\frac{10^{-14}}{4 \pi D^{2}} \int n_{\rm e} n_{\rm H} {\rm d}V$) & $(3.7^{+9.2}_{-1.8}) \times 10^{-3}$ & 
        $(1.56^{+0.91}_{-0.70}) \times 10^{-2}$ & $(3.1^{+3.2}_{-1.5}) \times 10^{-4}$ \\
$\Gamma$\tablenotemark{i} & & $2.79^{+0.11}_{-0.10}$ & 
        $2.85^{+0.10}_{-0.10}$ & $2.57^{+0.12}_{-0.11}$ \\
$K_{\rm pl, east}$\tablenotemark{j} & ($10^{-4}$ photons cm$^{-2}$ s$^{-1}$ keV$^{-1}$) & $5.95^{+0.71}_{-0.54}$ & 
        $6.41^{+0.54}_{-0.51}$ & $4.34^{+0.69}_{-0.62}$ \\
$K_{\rm pl, west}$\tablenotemark{k} & ($10^{-4}$ photons cm$^{-2}$ s$^{-1}$ keV$^{-1}$) & $12.6^{+1.7}_{-1.4}$ & 
        $13.5^{+1.6}_{-1.4}$ & $9.6^{+1.7}_{-1.1}$ \\
$\chi^{2}$ (dof) & & $459.8 (347)$ & $463.7 (347)$ & $430.6 (346)$ \\
Prob.\tablenotemark{l} & & $4.5 \times 10^{-5}$ & $2.7 \times 10^{-5}$ & $1.3 \times 10^{-3}$ \\

\enddata


\end{deluxetable}

\clearpage

\setlength{\parindent}{0pt}

$^{\rm a}$Model: B $=$ wind-blown bubble (model valid for gas
temperatures in the range from 8~eV to 80~keV); vB $=$ wind-blown
bubble with variable abundances (model valid for gas temperatures in
the range from 8~eV to 80~keV); M $=$ MEKAL (model valid for gas
temperatures in the range from 8~eV to 80~keV); P $=$ power-law. \\
$^{\rm b}$Measured column density towards the eastern part of the
bubble. \\
$^{\rm c}$Measured column density towards the western part of the
bubble. \\
$^{\rm d}$The gas temperature at the center of the bubble. \\
$^{\rm e}$C, N, Na, Al, Fe, and Ni abundance relative to solar of the
bubble. \\
$^{\rm f}$Fixed at the known value of $30$\% solar for the ISM in the LMC. \\
$^{\rm g}$O, Ne, Mg, Si, S, Ar, and Ca abundance relative to solar of
the bubble.  \\
$^{\rm h}$The normalization of the bubble model.  The units are
cgs. \\
$^{\rm i}$Photon spectral index. \\
$^{\rm j}$Normalization for the power-law model in the eastern part of
the bubble. \\
$^{\rm k}$Normalization for the power-law model in the western part of
the bubble. \\
$^{\rm l}$Probability that the model describes the data and that
$\chi^{2}$ exceeds the observed value by chance.\\

\clearpage

\clearpage

\clearpage

\clearpage

\clearpage

\end{document}